# Communication of information in systems of heterogeneous agents and systems' dynamics

Inga Ivanova[1]

**Abstract**

Communication of information in complex systems can be considered as major driver of systems evolution. What matters is not the communicated information by itself but rather the meaning that is supplied to the information. However informational exchange in a system of heterogeneous agents, which code and decode information with different meaning processing structures, is more complex than simple input-output model. The structural differences of coding and decoding algorithms in a system of three or more groups of agents, entertaining different sets of communication codes, provide a source of additional options which has an impact on system's dynamics. The mechanisms of meaning and information processing can be evaluated analytically in a model framework. The results show that model predictions accurately fit empirically observed data in systems of different origins.

**Key words:** Information, meaning, systems theory, autopoesis, model

## Introduction

[1] Institute for Statistical Studies and Economics of Knowledge, National Research University Higher School of Economics (NRU HSE), 20 Myasnitskaya St., Moscow, 101000, Russia

The functioning of information ecosystems depends on how people access, produce, consume and share information. This requires communication between human minds. Information is often addressed as pure messages or raw data which may entail grave misunderstanding with respect to the complexity of the process of inter-social communications. What really means is not the information by itself but the meaning which is attributed to information.

The theory of information ecosystem builds upon the notions of ecosystem and information. Ecosystem concept originates from ecology studies. From ecological perspective ecosystem can be defined as "*recycling flows of nutrients along pathways made up of living subsystems which are organized into process-orientated roles; connects living and non-living subsystems*" (Shaw and Allen, 2018, at p.90). There are three points to be noted in this definition: "*living subsystems*", "*flows of nutrients*", and "*process-orientated roles*". In a broader sense as applied to other domains this can be captured as "*agents*", "*interaction*", and "*heterogeneity*". With respect to social systems Kuehn (2023) proposed the definition of information ecosystem as "*all structures, entities, and agents related to the flow of semantic information relevant to a research domain, as well as the information itself.*" Thus ecosystem implies non-linear self-organizing system comprising of multiple agents, where system dynamics is a result of communication among agents.

Another attribute of information ecosystem is information. Generation and communication of information is the key driver of system's evolution. There is no agreed upon definition what does information mean. The notion of information is applied across wide range of disciplines so that it hardly be defined in transdisciplinary perspective (e.g. Floridi, 2011, Hofkirchner, 2013). Mingers and Standing (2018) structured existing theories of information with respect to different levels: empirical, syntactic, semantic, pragmatic, and social. At empirical level information is equated with data, at syntactic level it is counted as messages, semantic level includes the meaning of messages, pragmatic dimension considers the intentions of the sender and effects on receiver, social level regards the network of inter-social communications and selective mechanisms of individual cognitions. They further summarized the overall conception of information as "*one that leads us to see information systems as part of the wider human world of meaning processing through communications*" (2018, at p.14). From the general viewpoint information can be considered from the objective perspective as "environmental information", i.e. something that exists irrelevant to involvement of observer, or subjective perspective, as human or social relevant.

Bateson wishing to emphasize the meaning dimension of information defined information as "*difference which makes a difference*" (Bateson, 1972, at p.315). The messages by themselves do not make any difference to receiver. What makes information significant is the meaning attributed to messages (Yockey, 2005), which is generated with help of selection mechanisms (Luhmann, 1984).  Luhman (1996) argued that meaning is not intentionally produced by communicators but self-organizes in the process of communications like autopoetic system. Providing information with meaning forms the ground for further actions. Society's functional differentiation has an effect on communications, so that in a system with heterogeneous groups of agents the same information can be interpreted differently by members of each group and supplied with different meanings. For example, the ban on watching films in category 18+ by children is due to the fact that children may misinterpret the content of the film. Functional differentiation of different groups with respect to interpreting the information can be considered in the framework of symbolically generalized codes of communication (Parsons, 1968; Simon,

1973). Interaction among different sets of communication codes drives self-organization of meaning at the system level (Leydesdorff, 2021).

What is important about information is its measurement. Shannon (1948) developed a way to measure information in his mathematical theory of communications. He defined information as an uncertainty in terms of probabilistic entropy: $H = -\sum_i p_i log p_i$, so that the amount of information depends on the number of messages and their relative probabilities. Shannon abstained from discussing semantic aspects of communication as external to engineering problems. When information passes from the sender to receiver it can be disturbed by noise, generated due to technical reasons. This possibility is accounted for by introducing a box, labeled as "noise source" (Figure 1). Shannon's co-author Weaver proposed "semantic receiver" and "semantic noise" boxes as minor additions to diagram of a general communication system.

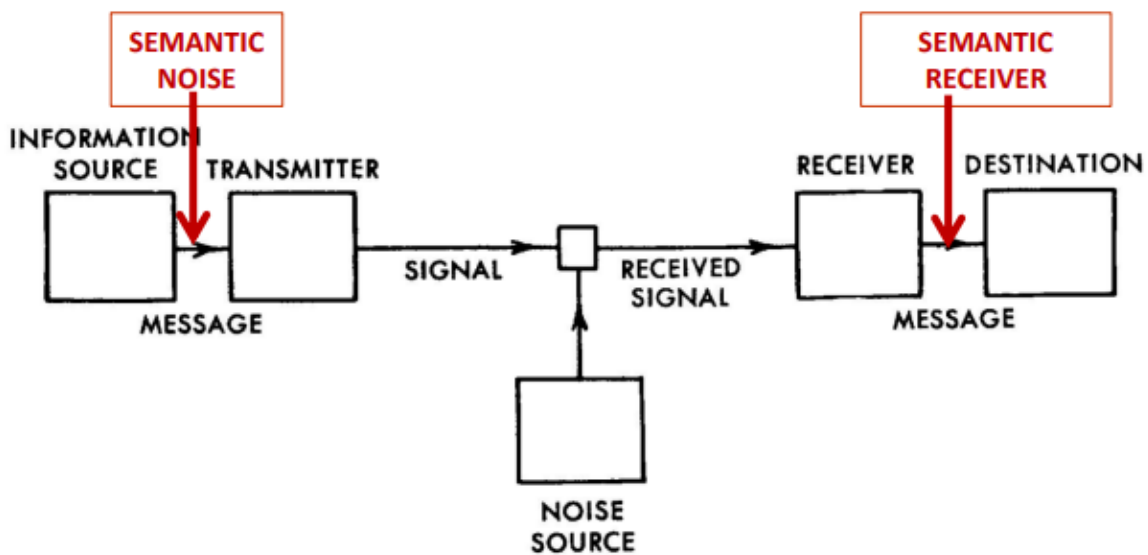

**Figure 1**: Schematic diagram of a general communication system. Source: Shannon (1948, at p. 380); with Weaver's box of "semantic noise" and a second source of "semantic noise" between the receiver and the destination.

These minor additions are intended to provide meaning to the information. Semantic noise encodes meaningful information into message and engineering transmitter transforms messages into signal. Engineering receiver changes signal into messages and semantic receiver subjects the messages to the second decoding to supply the messages with meanings. "Semantic noise" and "semantic receiver" encode and decode the information with help of some coding rules or communication codes. The sets of communication codes act as selection environments which organize different meanings into structural components (Leydesdorff, 2010). When information is communicated between agents, who use different communication code sets and provide different criteria to supply information with meaning, the same information can be differently processed by the sender and the receiver and supplied with different meanings. Communication codes may relate to individual (cultural) level, such as language, or to larger scale social systems with more specialized languages, as money, power or influence (Parsons, 1968).

Weaver (1949: 24) also suggested complement Shannon's diagram with two other levels that represent how meaning is conveyed and how it affect system dynamics, though he didn't specify the precise structure of these levels. Leydesdorff (2016) distinguished these levels in the following way ( Fig.2): information is communicated at the lower level of network relations, meaning is conveyed at the next level, sharing of meaning relies on correlations among generalized communication code sets, communication code sets, in turn, are subject to translations and reconstruction in process of communication as a response to their integration in instantiations.

.

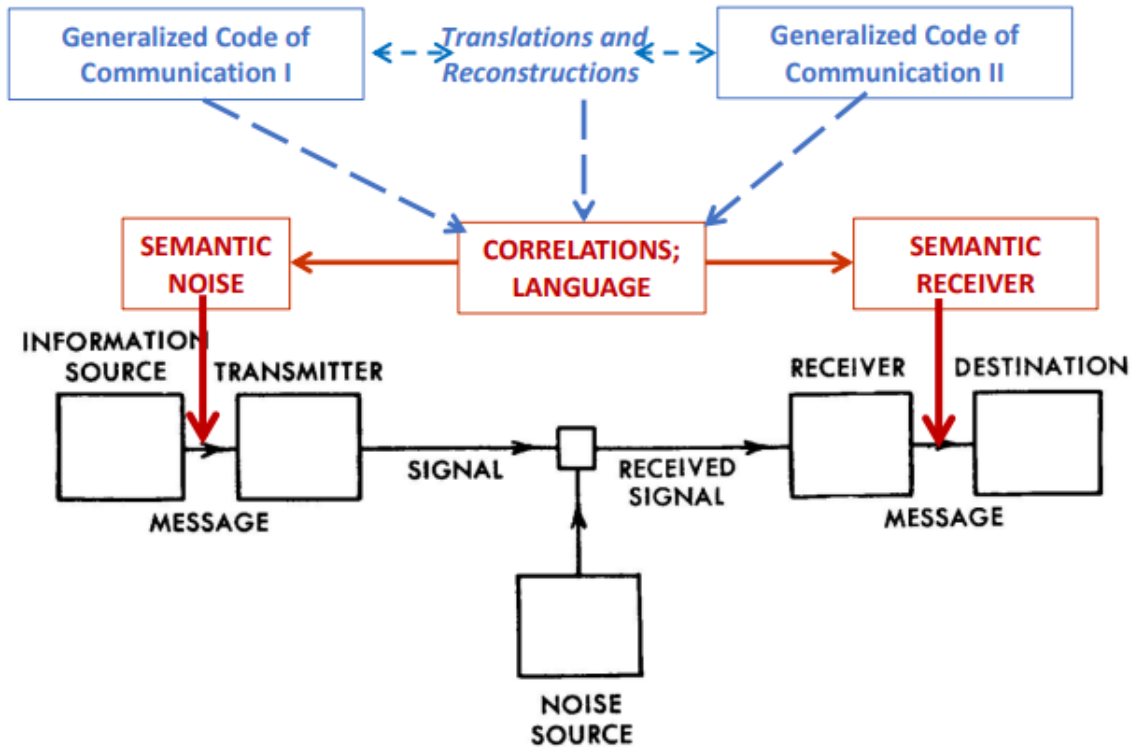

**Figure 2**: Additional levels added to the Shannon diagram (in red-brown and dark-blue, respectively) (source: Leydesdorff, 2016)

Meanings provide expectations about possible future states which are available, but not yet realized. These future states form a source of additional options for system evolution. Expectations operate as a feedback on the current state against the arrow of time. So that system simultaneously entertains its past, present and future states to steer its evolution. As the number of available options increase, so does the likelihood that the system will deviate from its current state. These additional options can be measured as redundancy which indicates the complement of information to the maximum informational content (Brooks & Wiley, 1986).

The task of developing sound strategies for improving information ecosystems raises questions about the real mechanisms that govern the evolution of systems, the knowledge of which can serve to predict the future states of systems. Schematic diagram in Fig.2 is far from measurement and operationalization. Operationalization of inter-social communications requires interdisciplinary approach, based on sociology, information theory, innovation studies, quantum physics, non-linear evolutionary equations, cybernetics, and complex systems theory.

### Mutual information, configurational information, and redundancy

Functionally differentiated social system can be mapped in a set theoretical representation. In case of two sets logical relation between two sets can be presented as Venn diagram in Fig. 3. Let each set consist of agents that have a common system of communication codes inherent in this set. Overlap comprises agents which incorporate communication codes from both sets. If the agents are provided with corresponding probabilities (e.g. with respect to possible options) one can operate these sets in terms of uncertainty distributions $H_1$ and $H_2$. Overlapping area $T_{12}$ is Shannon's mutual information.

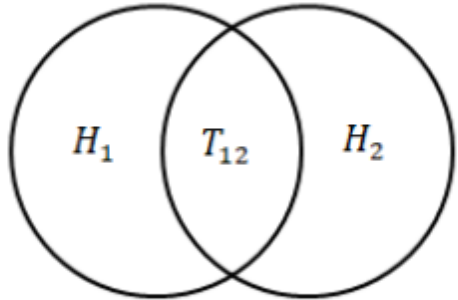

**Figure 3:** Set-theoretical representation of two overlapping distributions with informational contents $H_1$ and $H_2$

Total uncertainty $H_{12}$ corresponds to the surface square of Fig.3 which is the sum of overlapping uncertainties minus overlapping area:

$$H_{12} = H_1 + H_2 - T_{12} \tag{1}$$

Mutual information $T_{12}$ is positive (Theil, 1972: 59 ff.) and decreases total uncertainty.

For three overlapping distributions (Fig.4) last term in the formula right hand side, corresponding central overlap, changes the sign:

$$H_{123} = H_1 + H_2 + H_3 - T_{12} - T_{13} - T_{23} + T_{123} \tag{2}$$

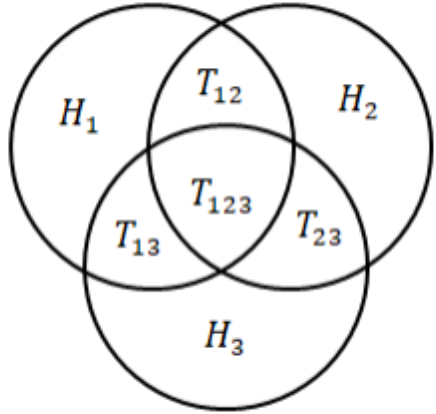

**Figure 4:** Set-theoretical representation of three overlapping distributions with information contents: $H_1, H_2, H_3$

Accordingly this term can't be mutual information in Shannon definition since it adds to the uncertainty and is referred to as configurational information (McGill, 1954). It can be shown that for higher dimensions the sign of last added term changes with each additional dimension so that it is signed information measure (Yeung, 2008).

This signed information measure can be considered as redundancy if one considers overlap as a surplus of uncertainty provided by additional options (Leydesdorff, Ivanova, 2014):

$$Y_{12} = H_1 + H_2 + T_{12} = H_{12} + 2T_{12} \tag{3}$$

Mutual redundancy in case of two dimensions equals mutual information with the opposite sign:

$$R_{12} = H_1 + H_2 - Y_{12} = -T_{12} \tag{4}$$

For higher dimensions it follows that mutual redundancy equals configurational information with repeatedly changed sign:

$$R_{123} = T_{123}$$

$$R_{1234} = -T_{1234}$$

When one measures configurational information one measures redundancy[2]. Processes of communicating information and meaning generation can happen at the same time withing the system, but mechanisms are different. Whereas information is transmitted via network of relations meaning can't be transmitted but is shared when communication codes sets overlap.

Mutual redundancy presents options which are available but not yet realized, i.e. it is a complement of existing uncertainty $H$ to maximum entropy $H_{max}$:

$$R = \frac{H_{max} - H}{H_{max}}$$

The measurement of mutual redundancy makes it possible to use it in empirical research. Redundancy was used as an indicator of synergy in the Triple Helix (TH) model of university-industry-government relations (e.g. Leydesdorff, 2003; Park & Leydesdorff, 2010; Leydesdorff & Strand, 2013 etc.).

### Triple Helix as inherently nonlinear social system

Triple Helix (TH) model of university-industry-government relations (Etzkowitz & Leydesdorff, 1995, 1998) can be consideres as the simplest example of non-linear self-organizing social system. Three larger groups of institutionally diggerentiated agents (otherwise called actors) present three interacting subdynamics forming overlay of communications (Etzkowitz &

---
[2] Krippendorff (1980, 2009a, 2009b) showed that Shannon-type information in multilateral interactions also exists and can be measured

Leydesdorff, 2000) which drive non-linear system evolution. Spheres of actors' activities are increasingly overlapping and actors can partially substitute for one another. Institutional differentiation entails also positional differentiation with respect to communication codes sets, which are correlated to some extent, i.e. meaning is provided from different perspectives, or positions (Burt, 1982). For example an invention can be considered with respect to its scientific significance, in terms of possible market entry and financial benefits, and from the point of view of legislative regulation. Non-linear TH dynamics is provided by system topology and can be modeled by non-linear equation. The same is true for Quadruple and higher order systems, while a system with two actors, such as e.g. a model of university-industry relations) remains linear and can be described by linear equation (Ivanova, Leydesdorff, 2014a).

When three or more actors communicate each third can disturb the communication between the other in reinforcing or stabilizing manner. Two cycles may emerge, as shown in Fig. 5. This mechanism (otherwise called as "triadic closure" is responsible for system formation (Bianconi et al., 2014; de Nooy & Leydesdorff, 2015)

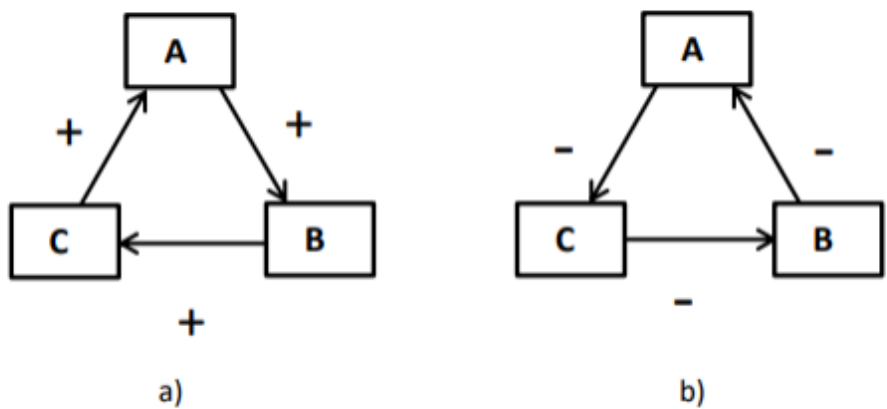

**Figure 5**: Schematic of three-component positive a) and negative b) cycles (Adapted from Ulanovitz, 2009)

Two cycles can simultaneously exist. Depending on the kind of cycle which prevail the system can either deviate from equilibrium or stabilize along historical trajectory. The trade-off between two dynamics can be presented in formula format as the difference between two three-dimensional rotating vectors *P* and *Q* (Ivanova & Leydesdorff, 2014b):

$$R(t) \sim P^2(t) - Q^2(t) \tag{5}$$

The first term in Eq. (5) adds to positive entropy and relates to historically realized options while the second term augments negative entropy and bears on new, not yet realized options. Historical realization refers to sequence of events which are generated via a recursive mode and self-organization relates to new, not yet realized options obtained through generation of meaning which is "*provided from the perspective of hindsight to events that have already happened or are happening ... codification operates hyper-incursively on meanings* " (Leydesdorff, 2021 at p. 154).[3]

**Evolutionary dynamics of expectations**

Redundancy can be considered as a proxy of additional options generated via biased processing of information. These options otherwise can be viewed as expectations. TH indicator allows

---
[3] Recursive system use their past states to modulate the present ones, incursive (or anticipatory) system employ past and present states to shape its present state, hyper-incursive system is reconstructed in terms of its future states (e.g. Rosen, 1985; Dubois, 1998; Leydesdorff & Dubois, 2004)

numerically capture the (static) value of these additional (unrealized) options in terms of probabilistic entropy. But what are the mechanisms that drive the dynamics of options?

Provided probabilistic entropy is subject to temporal cyclic fluctuations, corresponding probabilities $p_i$ can oscillate around their average values $p_{i0}$ in a harmonic mode (Dubois, 2019). When one of competing tendencies (historical realization or self-organization) prevails probabilities oscillate in a non-harmonic mode. In this case the probability density function $P$[4] satisfies the following non-linear evolutionary equation (Ivanova, 2022 a) and b)):

$$P_T - 6PP_X + P_{XXX} + C_1 = 0 \tag{6}$$

which is generalized form of Korteweg-de Vries (KdV) equation

$$U_T - UU_X + U_{XXX} = 0 \tag{7}$$

Eq. (7) has solutions in the form of solitary waves (solitons). A single soliton solution has the form:

$$P(X,T) = 2\left(\frac{\kappa}{2}\right)^2 ch^{-2}\left[\frac{\kappa}{2}\left(X - 4\left(\frac{\kappa}{2}\right)^2 T + \frac{C_1}{2}T^2\right)\right] - C_1 T \tag{8}$$

In general, the $N$ – soliton solution of Eq. (6) can be written as:

$$P = 2\frac{d^2}{dX^2}\log F_N \tag{9}$$

where:

$$F_N = exp\left[-\frac{C}{2}tx^2 + Ax + B\right] \cdot \sum_{\mu=0,1} exp\left(\sum_{i=1}^{N} \mu_i \eta_i + \sum_{1 \le i < j}^{N} \mu_i \mu_j A_{ij}\right) \tag{10}$$

---

[4] Probability $p$ and probability density $P$ are related as following: $p(t) = \int_{-\infty}^{t} P(\tau)d\tau$

Here: $\eta_i = k_i X - k_i^3 T$; $A_{ij}$ are the phase shifts of the solitons: $e^{A_{ij}} = \left(\frac{k_i - k_j}{k_i + k_j}\right)^2$ (cf. Ablowitz and Segur, 1981)[5]. Initial irritation, depending on its size develops in oscillations, single solitary wave, or a train of such waves (Lax, 1974). In case of a sequence of waves originating from single irritation there is a relationship between soliton amplitudes and time intervals:

$$\frac{A_i}{A_j} = \frac{T_i}{T_j} \quad \text{or} \quad \frac{A_i - A_j}{T_i - T_j} = \text{const} \tag{11}$$

Information obtained via informational exchange is processed with communication codes and turned into expectations, which are generated at a system's level with respect to future time. Here expectations are analytical events (options) and actions are historical events,[6] which can be observed over some time as a response to the expectations (as if expectations move against the arrow of time and turn into actions). There is a dynamics of the actions in historical perspective at the system's bottom level and a dynamic of expectations at the upper level. Expectations are eventually transformed into actions and represent new system states.[7]

Initial expectations (irritation) are projected to the future and further develop in waves following the non-linear dynamics of information processing. Finally expectations are realized and turned into observed system changes forming specific wave patterns in t-axes.

---

[5] The sum over $\mu = 0,1$ refers to each of $\mu_i$. E.g. performing the calculation for N=3 yields $F_3 = 1 + e^{\eta_1} + e^{\eta_2} + e^{\eta_3} + e^{\eta_1 + \eta_2 + A_{12}} + e^{\eta_1 + \eta_3 + A_{13}} + e^{\eta_2 + \eta_3 + A_{23}} + e^{\eta_1 + \eta_2 + \eta_3 + A_{12} + A_{13} + A_{23}}$

[6] Shannon (1948) defined the proportion of non-realized but possible options as redundancy, and the proportion of realized options as the relative uncertainty or information.

[7] According to the second law of thermodynamics a system's entropy increase with time. For isolated systems it can reach thermodynamic equilibrium

The described mechanism operates at different time scales generating a self-similar fractal structure and reveals underlying dynamics of codes of communication. The next paragraph shows some examples of this dynamics.

## Applications

The feasibility of approach for real data analysis can be illustrated by three unrelated cases: infectious disease spread, investors' behavior in financial markets, and rumors propagation. The selection is based upon topological and functional similarity of systems under consideration. In all cases one can differentiate the typical three positionally differentiated groups which interact upon communicating of information, where information appears in different forms.

### a) Infectious disease spread (the case of Covid-19)

One of popular approaches in infectious disease dynamics studies relies on compartment models which consider the population assigned to three or more groups or compartments. Kermack and McKendrick, (1927) proposed a model, otherwise called *SIR* model, where population is subdivided into three groups – people susceptible to infection (*S*), infected (*I*), and individuals that are recovered or removed from the total population (*R*). The model comprises the system of ordinary differential equation connecting compartments' representing variables. Provided initial condition *S(0)*=1 the following conservation law holds: *S(t)+I(t)+R(t)*=1, i.e. the total population remains constant. *S* and *I* variables are directly coupled and variable *R* can impact the previous two through the conservation law. The removed individuals can be further transformed into susceptible ones due to expiration of immunity or mutation of the infectious agent, so that triadic closure $S \to I \to R \to S$ as shown in Fig. 5 is formed. In this respect *SIR* model can be compared to TH model where infection plays the role of information which is transmitted between compartments causing different reactions from recipients due to different immune status of groups' members.

*SIR* model system of equations can be sequentially reduced to Verhulst equation which has the solution in the form of logistic function (Postnikov, 2020). Logistic functions are used to fit cumulative number of (e.g. infected) cases. Derivative of logistic function correspond single soliton solution of KdV equation (Eq. 7). In practice empirical data can be rarely fitted by single function and one often has to use multi-logistic approximation (e.g. Rządkowski & Figlia, 2021, Postnikov, 2021). Derivative of multi-logistic function corresponds to multi-soliton solution of KdV equation.

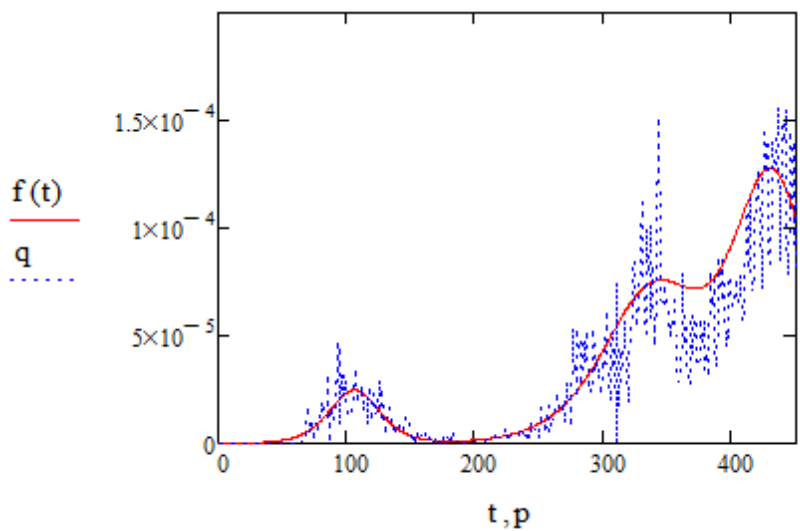

**Figure 6**. Time series Finland daily new Covid-19 cases and its three wave approximation, x axis: days since the first case, y axis: dot line - empirical data, solid line – function fit (source: Ivanova, 2022)

Fig.6 presents a graph of Finland daily new Covid-19 cases and model predicted values obtained via multi-soliton approximation. Corresponding amplitude and time ratios are: $\frac{A_2}{A_1} = 2.83$; $\frac{A_3}{A_1} = 4.87$; $\frac{T_2}{T_1} = 3.3$; $\frac{T_3}{T_1} = 4.3$ which approximately corresponds the relation given by Eq. (11) (Ivanova, 2022a). The observed wave train is a result of single initial perturbation and can be attributed to single infection wave. The results suggest comparatively good agreement between empirically observed data and model predicted values.

### b) Financial markets

Financial market dynamics is represented by the interplay among investors entertaining different expectations about future asset price movement. In this respect investors can be subdivided into three groups– those who prefer to buy asset, sell asset, and those who temporally abstain from active trading. Their decisions are based on available market information which is differently processed by members of each group according different vision of future market situation (in other words they entertain different sets of communication codes).

It follows from the model that when one or other tendency prevails (Eq. 5) asset prices develop in trends which can be described by a non-linear evolutionary equation (Eq. 6). Though financial markets demonstrate very complex dynamics, some trends can be fitted by train of solitons. An example of such behavior is presented in Fig. 7.

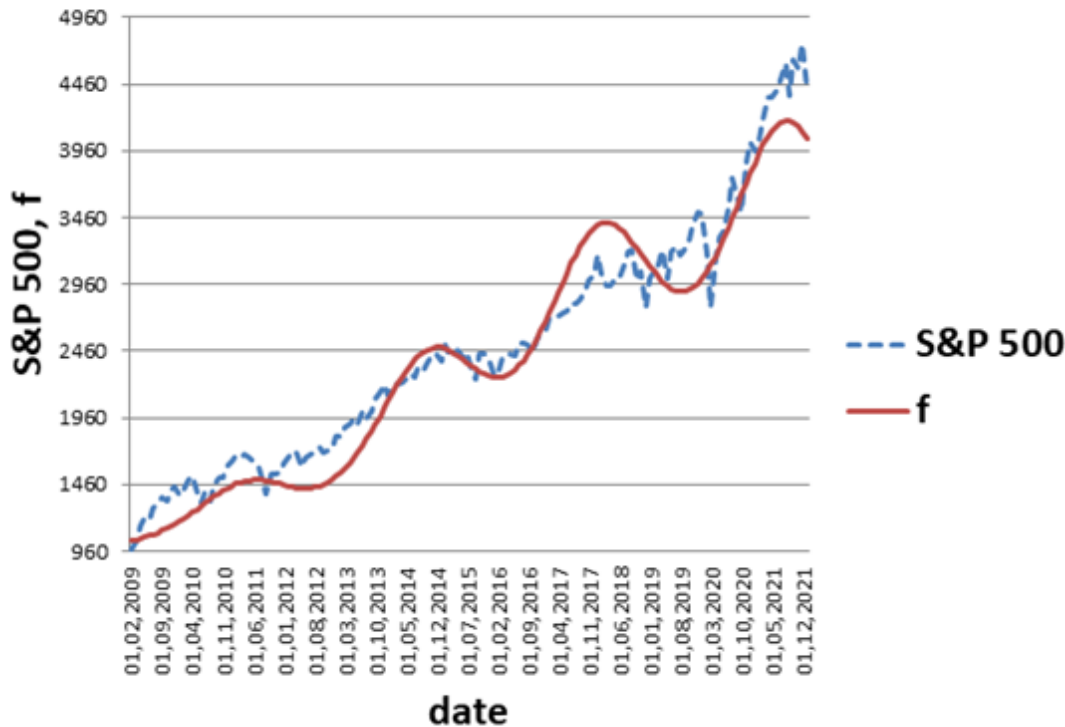

**Figure 7:** Time series for monthly Standard and Poor's 500 stock market index (S&P500) empirical data and model (*f*) fit (source: Ivanova, 2022b).

Fig. 7 is a chart for Standard and Poor's 500 stock market index (S&P500) historical monthly inflation adjusted data for the period: 2009 – 2022 and their model approximation (*f*). The data are fitted with a series of four waves. Successive amplitude to time ratios of each two adjacent waves ($\frac{A_i - A_j}{T_i - T_j}$): 19.74; 20.29; 20.91 almost do not change which suits the model predictions. R squared value for OLS regression between empirical data and model fit data is 0.95. Engel - Granger two-step co-integration test (Engel & Granger, 2015) eliminates the problem of spurious correlation between observed asset's price and model predicted curve. Null hypothesis of a unit root can be rejected at the 5% level (Ivanova, 2022b).

Presented model can be used to predict the trend reversal. When the trend starts to unfold and wave structure is confirmed one can await certain price levels to be reached at certain time points. If the level is reached by time scheduled point one can expect (at least temporal) trend reversion. In presented case of SP 500 the model suggests that one should expect a trend reversal, which corresponds to the beginning of a recession period.

However the presented model should not be considered a universal tool for forecasting financial time series during any periods of trend and price consolidation given that Equation 7 is derived under certain conditions that may not always be met.

### c) Rumors propagation

Daley and Kendall (1964, 1965) had drawn a parallel between infectious disease spread and rumor propagation. They studied the problem from the viewpoint of mathematical epidemiology and formulated rumor propagation model in analogy with *SIR* model by subdividing population into three groups: *I* – the fraction of individuals who has not heard rumor; *S* – people actively spreading rumor; *R* – those who no longer spreading rumor (stiflers). These groups are respectively identical to *SIR* model groups: susceptible (*S*), infected (*I*), and recovered/removed (*R*). The rumor is spread via contacts between members of *S* and *I* groups.

The difference with *SIR* model lies in the mechanism of rumor decay. When a spreader contacts another spreader or stifler the initiating spreader becomes a stifler. Due to forgetting process or rumor change stiflers can become ignorant. A set of three differential equations governs the system dynamics. The model with various modifications with respect to the number of compartments and interaction mechanisms was extensively used for the description of rumor propagation in social networks, (e.g. Yu *et al*., 2021, Chang, 2023), computer viruses propagation (e.g. Mishra and Saini, 2007, Piqueira and Araujo, 2008), and effect of unexpected events on market dynamics (McDonald *et al*., 2008). Goffman and Newill (1964) considered an application of compartment epidemic theory to transmission of ideas in science, religious fields, etc.

Rumors reflect as the interest of people in certain information. The number of requests in internet search engines can be considered as a proxy of such interest. Fig.8 presents the results of daily query by word "bitcoin" in Google Trends (https://trends.google.com) for period: 19.03.2019-01.08.2019.

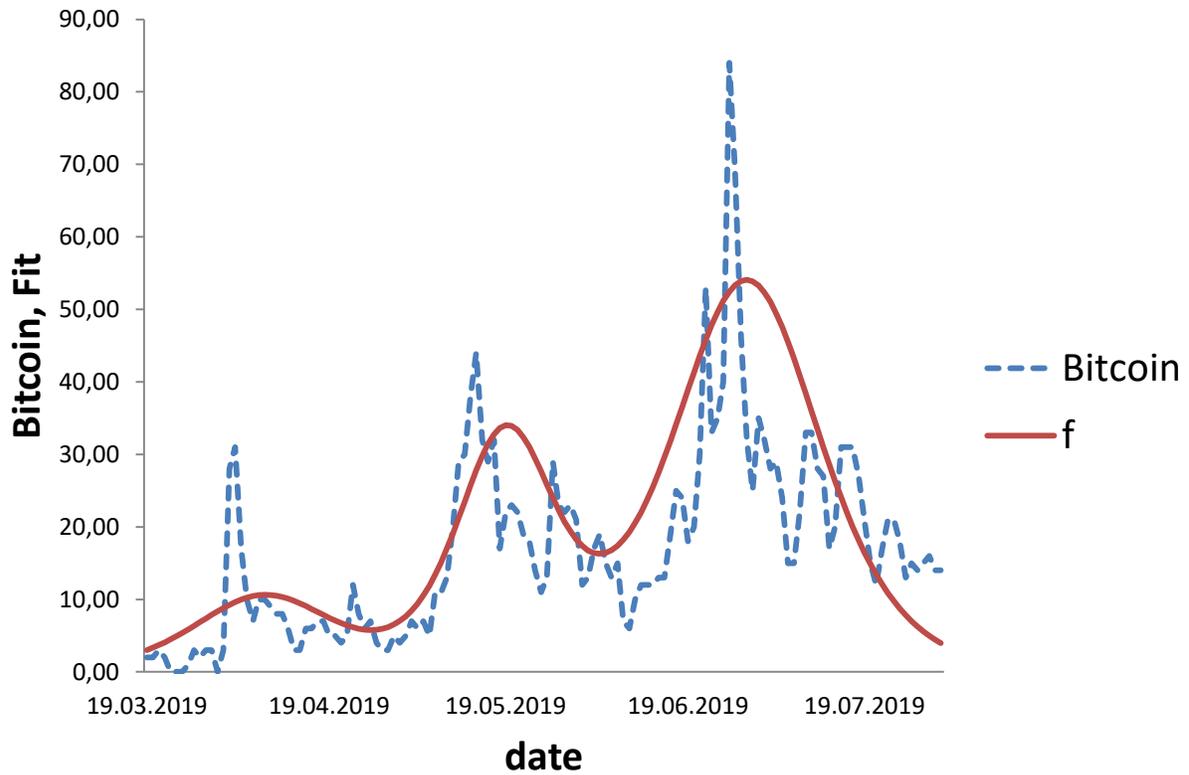

**Figure 8:** Daily query by word "bitcoin" in Google Trends for period: 19.03.2019-01.08.2019, empirical data and model (*f*) fit

Empirical data are fitted into the three wave curve where each wave has characteristic shape of the solitary wave:

$$f(t) = \sum_{i=1}^{3} A_i ch^{-2}[B_i(t - T_i)]$$

Parameters of approximation are: $A_1 = 10.54$; $A_2 = 32.15$; $A_3 = 53.76$; $B_1 = 0.062$; $B_2 = 0.09$; $B_3 = 0.06$; $T_1 = 20$; $T_2 = 61$; $T_3 = 102$. The following relation between amplitudes ratios and time shift ratios holds: $A_2/A_1 = T_2/T_1 = 3.05$; $A_3/A_2 = T_3/T_2 = 1.67$. There is a linear correlation between empirical and model estimated data, Pearson correlation coefficient value is 0.745.

Spurious correlation between empirical an model fit data can be eliminated with help of Engel - Granger two-step co-integration test. Dickey–Fuller t-distribution critical values for N=250 without trend are -3,46 for 1% level and -2.88 for 5% level (Fuller, 1976, p.373), computed value equals – 5.14. Thus one can conclude that a null hypothesis of a unit root can be rejected at the at the 1% level.

## Discussion, conclusion

In his study of communicational phenomena in organizations german sociologist Niklas Luhmann (2018) considered organizations as substantially the product of communicative activities. He ultimately stated that it is not human who communicate, but "only communications can communicate" (Luhmann, 2002 at p. 169), implying that human can introduce information into the system but then information develops according to system's specific information processing operation which can't be controlled from outside. Externally communicated information produces "irritations" which induce further communications determined by self-organizing operating mechanisms. However Luhmann's communication approach remained only qualitative.

Drawing on Luhman's take on organizations as autopoetic communication systems, the concept of anticipatory systems (Rosen, 1985, Dubois, 2000 and 2003), and Shannon's information theory Leydesdorff (2021) suggested a theory of discursive communications with an accent on the concept of meaning. According this theory information in social communications is processed via specific sets of communication codes which span horizons of meaning acting as selection and coordination mechanisms. Meanings generate expectations and incurred on events against the arrow of time. Different code sets shape each other during communication.[8] Expectations can be measured using entropy calculation as redundancy (i.e. additional options) with help of Shannon's information theory. He also incorporated the conceptual framework of the Triple Helix model of university-industry-government relations (Etzkowitz & Leydesdorff, 1995, 1998) to develop better metrics for socio-economic effectiveness by considering on redundancy in the system as a measure of synergy, that formed the basis for numerous theoretical and empirical works elaborating on quantitative estimation of information and meaning static and dynamics characteristics in complex systems of different origins (Leydesdorff, 2008; Leydesdorff & Dubois, 2004; Leydesdorff, Dolfsma, Van der Panne, 2006; Leydesdorff & Franse, 2009; Leydesdorff & Ivanova, 2014; Leydesdorff, Petersen & Ivanova, 2017; Ivanova (2022 a) and b)).

The present paper is a further development of quantitative theory of information and meaning with potentially numerous applications. The theory main points are as follows: a system can be considered as a network of relations among three groups of positionally differentiated agents because three is the minimal number of dimensions where the complex dynamics emerge. The number of groups can be enlarged in higher topologies, but higher topologies are decomposable into triads (Freeman, 1996). System's dynamics is analyzed from an information theory perspective taking into account the relationships between information processing and meaning generation within the system. Information is communicated via the network of relations.

---

[8] In Luhmann's theory there is no interactions among codes since subsystems are set as "operationally closed"

Meaning generation takes place at the systemic level on the top of this network. Different agents use different communication codes, reflecting their preferences, to provide meaning to the information. Codes can be considered the eigenvectors in a vector-space (von Foerster, 1960) and structure the communications as selection environments. Communicated information is supplied with different meanings by different agents. Meaning is provided from the perspective of hindsight. Meanings cannot be communicated, as in case of information, but only shared. Providing information with meaning increases the number of options (redundancy). This mechanism can be considered probabilistically using Shannon's equations (Shannon, 1948). The generation of options (redundancy) is crucial for system change. The trade-off between the evolutionary generation of redundancy and the historical variation providing uncertainty can be measured as negative and positive information, respectively. The dynamics of information, meaning, and redundancy can be evaluated empirically using the sign of mutual information as an indicator. When the dynamics of expectations, generating redundancies, prevail over the historical construction generating entropy, mutual redundancy is negatively signed because the relative uncertainty is reduced by increasing the redundancy. The balance between redundancy and entropy can be mapped in terms of two vectors (*P* and *Q*) which can also be understood in terms of the generation versus reduction of uncertainty in the communication that results from interactions among the three (bi-lateral) communication channels.

The feedback loops of communication shape and reproduce information in the form that was neither initially intended nor wholly understandable by system's agents, according self-organizing mechanisms, and feed the system dynamics. These mechanisms can be captured analytically in a model representation by non-linear evolutionary equation.

The results obtained in the framework of information approach are supported by findings of other specific case studies. Paxson and Shen (2022) considered of infectious diseases spread in the framework of *SIR* model. They showed that under the weekly non-linear assumption *SIR* model can be reduced to KdV equation. Jenks (2020) showed that the behavior of financial markets in periods of crises can be described by rogue waves. This type of waves is reported analytically in the nonlinear option pricing model (Yan, 2010). Rogue waves can be found in Korteweg de-Vries (KdV) systems if real nonintegrable effects, higher order nonlinearity and nonlinear diffusion are considered (Lou & Lin, 2018). Dhesi & Ausloos (2016) when studying agent behaviour reacting to time dependent news on the log returns in the framework of Irrational Fractional Brownian Motion model observed a kink-like effect reminiscent of soliton behavior. They further posed a question - what is the differential equation whose solution describes this effect? And KdV equation appears to be the best candidate for this position. Although mentioned case studies do not address the underlying drivers behind the non-linear phenomena described. In this regard, the information approach seems to be a universal tool for studying system dynamics from the point of view of information transfer for a wide range of systems of various origins.

Communication of information approach to system dynamics has great potential to expand our capabilities in understanding the dynamics of complex systems and improve our forecast capabilities concerning events that have yet to take place. It can prove to be useful when applied to different domains connected with informational exchange in composite systems. Maybe even more important is the idea that meaning can be measured quantitatively since when there is a measure, one also can dose. The model suggests that information ecosystems are sensitive not only to the synergy of agents' interaction but also to the amount and timing of external news arrival. The same information being introduced sequentially cause different system's reaction than if being introduces simultaneously.

The complex process of information flow in ecosystems can be reduced to its essential properties. Understanding basic mechanisms which control information flow can help policymakers and practitioners in various domains in their everyday activity to design the most appropriate and effective strategies to achieve their goals by not only reacting to the environment but shaping this environment.

At the same time the paper adds to a growing literature that studies the role of social norms, moral attitudes, religions and ideologies in imperfect rationality in inter-social communications.

**References**

Ablowitz, M. J. and Segur, H. (1981). *Solitons and the Inverse Scattering Transform*. Philadelphia: SIAM. https://doi.org/10.1137/1.9781611970883

Bateson, G. (1972). *Steps to an Ecology of Mind*. New York: Ballantine.

Bianconi, G., Darst, R. K., Iacovacci, J., & Fortunato, S. (2014). Triadic closure as a basic generating mechanism of communities in complex networks. *Physical Review E, 90*(4), 042806.

Brooks, D. R., & Wiley, E. O. (1986). *Evolution as Entropy*. Chicago/London: University of Chicago Press.

Burnside, C., Eichenbaum, M., Rebelo, S. T. (2016). Understanding booms and busts in housing markets. *J. Polit. Econ. 124*, 1088–1147.

Burt, R. S. (1982). *Toward a Structural Theory of Action*. New York, etc.: Academic Press.

Chang, X. (2023). Study on an SIR rumor propagation model with an interaction mechanism on WeChat networks. *Front. Phys., Sec. Social Physics*, 10 - 2022 https://doi.org/10.3389/fphy.2022.1089536

Daley, D.J. and Kendall, D. G. (1964). Epidemics and Rumours. *Nature 204*, 1118

Daley, D.J., and Kendal, D.G. (1965). Stochastic rumors. *J. Inst. Maths Applics 1*, p. 42.

Dhesi, G., & Ausloos, M. (2016). Modelling and measuring the irrational behaviour of agents in financial markets: Discovering the psychological soliton. *Chaos, Solitons & Fractals, 88*, 119-125.

Dubois, D. M. (2000). Review of Incursive, Hyperincursive and Anticipatory Systems – Foundation of Anticipation in Electromagnetism. In D. M. Dubois (Ed.), *Computing Anticipatory Systems CASYS'99* (Vol. 517, pp. 3–30). Liege: Amercian Institute of Physics.

Dubois, D. M. (2003). Mathematical foundations of discrete and functional systems with strong and weak anticipations. In M. V. Butz, O. Sigaud, & P. Gérard (Eds.), *Anticipatory behavior in adaptive learning systems (lecture notes in artificial intelligence* (Vol. 2684, pp. 110–132). Berlin: Springer.

Dubois D.M. (2019). Generalization of the discrete and continuous Shannon entropy by positive definite functions related to the constant of motion of the non-linear Lotka-Volterra system. *J. Phys.: Conf. Ser.* 1251 012012

Duffy J., Rabanal J.P and Rud O.A. (2021). The impact of ETFs in secondary asset markets: Experimental evidence. *Journal of Economic Behavior and Organization 188* (2021) 674-696, https://doi.org/10.1016/j.jebo.2021.06.003

Engle, R. F. and Granger, C, W. J. (2015). Co-Integration and Error Correction: Representation, Estimation, and Testing. *Applied Econometrics, 39*(3), 107–135.

Etzkowitz, H. & Leydesdorff, L. (1995). The Triple Helix—University-Industry-Government Relations: A Laboratory for Knowledge-Based Economic Development. *EASST Review 14*(1), 14-19.

Etzkowitz, H. & Leydesdorff, L. (1998). The endless transition: A "triple helix" of university – industry – government relations, *Minerva 36*, 203-208.

Etzkowitz, H., & Leydesdorff, L. (2000). The Dynamics of Innovation: From National Systems and 'Mode 2' to a Triple Helix of University-Industry-Government Relations. *research Policy, 29*(2), 109-123.

Floridi, L. (2011). The Philosophy of Information. Oxford: Oxford University Press.

Freeman, L. C. (1996). Cliques, Galois lattices, and the structure of human social groups. Social Networks, 18(3), 173–187.

Fuller, W. A. (1976). *Introduction to Statistical Time Series.* New York: John Wiley & Sons.

Goffman, W. and Newill, V. A. (1964). Generalization of epidemic theory: an application to the transmission of ideas. *Nature, 204* (4955), 225–228.

Hirshleifer, D. (2020). Presidential address: Social transmission bias in economics and finance. *J. Finance 75*, 1779–1831.

Hofkirchner, W. (2013). Emergent Information: A Unified Theory of Information Framework. Singapore: World Scientific Publishing

Ivanova, I. (2022a). Information Exchange, Meaning and Redundancy Generation in Anticipatory Systems: Self-organization of Expectations - the case of Covid-19. *International Journal of General Systems, 51*(7), 675-690. doi: 10.1080/03081079.2022.2084727

Ivanova, I. (2022b). Evolutionary Dynamics of Investors Expectations and Market Price Movement. Available online: https://arxiv.org/abs/1912.11216. doi: 10.48550/arXiv.1912.11216

Ivanova, I. and Leydesdorff, L. (2014a). Rotational Symmetry and the Transformation of Innovation Systems in a Triple Helix of University-Industry-Government Relations. *Technological Forecasting and Social Change 86,* 143-156. doi: 10.1007/s11192-014-1241-7

Ivanova, I. and Leydesdorff, L. (2014b). A simulation model of the Triple Helix of university-industry-government relations and the decomposition of the redundancy. *Scientometrics, 99*(3), 927-948. doi: 10.1007/s11192-014-1241-7

Jenks, T. (2020). *Hyperwave Theory: The Rogue Waves of Financial Markets*. Archway Publishing.

Kermack, W.O., McKendrick, A.G. (1927). A contribution to the mathematical theory of epidemics. *Proc. R. Soc. Lond. A, 115*,700–21. doi:10.1098/rspa.1927.0118

Kuehn, E. (2023). The information ecosystem concept in information literacy: A theoretical approach and definition. *JASIST*, *4* (3), 434-443

Krippendorff, K. (1980). Q; an interpretation of the information theoretical Q-measures. In R. Trappl, G. J. Klir & F. Pichler (Eds.), Progress in cybernetics and systems research (Vol. VIII, pp. 63-67). New York: Hemisphere.

Krippendorff, K. (2009a). W. Ross Ashby's information theory: a bit of history, some solutions to problems, and what we face today. International Journal of General Systems, 38(2), 189-212.

Krippendorff, K. (2009b). Information of Interactions in Complex Systems. International Journal of General Systems, 38(6), 669-680.

Lax P. (1974). Periodic Solutions of the KdV equations. *Lectures in Applied Mathematics 15*, 85-96. https://doi.org/10.1002/cpa.3160280105

Leydesdorff, L. (2003). The Mutual Information of University-Industry-Government Relations: An Indicator of the Triple Helix Dynamics. *Scientometrics, 58*(2), 445-467.

Leydesdorff, L. (2008). The Communication of Meaning in Anticipatory Systems: A Simulation Study of the Dynamics of Intentionality in Social Interactions. In D. M. Dubois ed.,

Leydesdorff, L. (2010). The Knowledge-Based Economy and the Triple Helix Model. *Annual Review of Information Science and Technology, 44*, 367-417

Leydesdorff, L. (2016). Information, meaning, and intellectual organization in networks of interhuman communication, pp. 280–303 in: Cassidy R. Sugimoto (Ed.), Theories of informetrics and scholarly communication: A festschrift in honor of Blaise Cronin, . In C. R. Sugimoto (Ed.), *Theories of informetrics and scholarly communication: A festschrift in honor of Blaise Cronin*. Berlin/Boston MA: De Gruyter.

Leydesdorff, L. (2021). The Evolutionary Dynamics of Discursive Knowledge: Communication-Theoretical Perspectives on an Empirical Philosophy of Science. In: *Qualitative and Quantitative Analysis of Scientific and Scholarly Communication* (Wolfgang Glänzel and Andrasz Schubert, eds., Cham, Switzerland: Springer Nature.

Leydesdorff, L., Dolfsma, W., & Van der Panne, G. (2006). Measuring the knowledge base of an economy in terms of triple-helix relations among 'technology, organization, and territory'. *Research Policy, 35*(2), 181-199. doi: 10.1016/j.respol.2005.09.001

Leydesdorff, L., & Dubois, D. M. (2004). Anticipation in Social Systems: The Incursion and Communication of Meaning. *International Journal of Computing Anticipatory Systems, 15*, 203-216.

Leydesdorff, L., & Franse, S. (2009). The Communication of Meaning in Social Systems. *Systems Research and Behavioral Science, 26*(1), 109-117. doi: 10.1002/sres.921

Leydesdorff, L., & Ivanova, I. A. (2014). Mutual Redundancies in Interhuman Communication Systems: Steps Toward a Calculus of Processing Meaning. *Journal of the Association for Information Science and Technology, 65*(2), 386-399. doi: 10.1002/asi.22973

Leydesdorff, L., Petersen, A., & Ivanova, I. (2017). Self-Organization of Meaning and the Reflexive Communication of Information. *Social Science Information 56*(1), 4-27; doi: 10.1177/0539018416675074

Leydesdorff, L., & Strand, O. (2013). The Swedish system of innovation: Regional synergies in a knowledge-based economy. *Journal of the American Society for Information Science and Technology, 64*(9), 1890-1902. doi: 10.1002/asi.22895

Lou, S., Lin, J. (2018). Rogue Waves in Nonintegrable KdV-Type Systems. *Chinese Physical Letters, 35*(5), 050202

Luhmann, N. (1984). *Soziale Systeme. Grundriß einer allgemeinen Theorie*. Frankfurt a. M.: Suhrkamp

Luhmann, N. (2002). How Can the Mind Participate in Communication? in N. Luhmann, *Theories of Distinction: Redescribing the Descriptions of Modernity*. Stanford, CA: Stanford University Press, 169–186

*Proceedings of the 8th Intern. Conf. on Computing Anticipatory Systems CASYS'07* (Vol. 1051 pp. 33-49). Melville, NY: American Institute of Physics Conference Proceedings.

Luhmann, N. (2018). *Organization and decision*. Cambridge, UK: Cambridge University Press.

McDonald, M., Suleman,O., Williams, S., Howison, S., and Johnson, N. F. (2008). Impact of unexpected events, shocking news, and rumors on foreign exchange market dynamics. *Physical Review E, 77*(4). Article ID 046110

Mingers, J., and Standing, G. (2018). What is information? Toward a theory of information as objective and veridical. *Journal of Information Technology, 33*(2), 85-104. doi: 10.1057/s41265-017-0038-6

Mishra, B. K. and Saini, D. (2007). Mathematical models on computer viruses. *Applied Mathematics and Computation*, *18* (2), 929–936.

McGill, W.J. (1954). Multivariate information transmission. *Psychometrika, 19*, 97-116.

de Nooy, W., & Leydesdorff, L. (2015). The dynamics of triads in aggregated journal–journal citation relations: Specialty developments at the above-journal level. Journal of Informetrics, 9(3), 542–554. https://doi.org/10.1016/j.joi.2015.04.005

Park, H. W., & Leydesdorff, L. (2010). Longitudinal trends in networks of university-industry-government relations in South Korea: The role of programmatic incentives. *Research Policy, 39*(5), 640-649.

Parsons, T. (1968). Interaction: I. Social Interaction. In D. L. Sills (Ed.), *The International Encyclopedia of the Social Sciences* (*Vol. 7*, pp. 429-441). New York: McGraw-Hill.

Paxson, W., Shen, B.-W. (2022). A KdV–SIR Equation and Its Analytical Solutions for Solitary Epidemic Waves. *International Journal of Bifurcation and Chaos, 32* (13), 2250199. doi:10.1142/S0218127422501991

Piqueira, J. R. C. and Araujo, V. O. (2008). A modified epidemiological model for computer viruses. *Applied Mathematics and Computation, 213* (2), 355–360.

Postnikov, E. (2020). Estimation of Covid-19 dynamics "on a back of envelope": Does the simplest SIR model provide quantitative parameters and predictions? *Chaos, Solitons and Fractals, 135*, 109841. doi: 10.1016/j.chaos.2020.109841

Postnikov, E. B. (2021). Reproducing country-wide COVID-19 dynamics can require the usage of a set of SIR systems. *Peer J*, 9, e10679. doi: peerj.com/articles/10679

Rosen, R. (1985). *Anticipatory Systems: Philosophical, Mathematical and Methodological Foundations*. Oxford, etc.: Pergamon Press.

Rządkowski, G., & Figlia, G. (2021). Logistic Wavelets and Their Application to Model the Spread of COVID-19 Pandemic. *Applied Sciences, 11*, 8147.

Shannon, C. E. (1948). A Mathematical Theory of Communication. *Bell System Technical Journal, 27*, 379-423 and 623-656.

Shaw, D., Allen, T. (2018). Studying innovation ecosystems using ecology theory. *Technol. Forecast. Soc. Chang. 136*, 88–10

Shiller, R. J. (2019). *Narrative economics: How Stories Go Viral and Drive Major Economic Events*. Princeton, NJ: Princeton University Press.

Simon, H. A. (1973). The Organization of Complex Systems. In H. H. Pattee (Ed.), *Hierarchy Theory: The Challenge of Complex Systems* (pp. 1-27). New York: George Braziller Inc.

Theil, H. (1972). Statistical Decomposition Analysis. Amsterdam/ London: North-Holland.

Weaver, W. (1949). Some Recent Contributions to the Mathematical Theory of Communication. In C. E. Shannon & W. Weaver (Eds.), *The Mathematical Theory of Communication* (pp. 93-117.). Urbana: University of Illinois Press.

Ulanowicz, R. E. (2009). The dual nature of ecosystem dynamics. *Ecological modelling 220*(16), 1886-1892.

von Foerster, H. On self-organizing systems and their environments. In: *Self-Organizing Systems*, M.C. Yovits, S. Cameron eds. (1960). Pergamon, London, pp. 31–50

Yan, Z. (2010). Financial Rogue Waves. *Communications in Theoretical Physics, 54*(4), 947-949. doi:10.1088/0253-6102/54/5/31

Yeung, R. W. (2008). *Information Theory and Network Coding*. New York, NY: Springer

Yockey, H.P. (2005). *Information Theory, Evolution, and the Origin of Life*. Cambridge University Press.

Yu, Z., Lu, S., Wang, D., Li, Z. (2021). Modeling and analysis of rumor propagation in social networks. *Information Sciences, 580*, 857-873.